\documentclass[journal]{IEEEtran}
\usepackage{amssymb}
\usepackage[cmex10]{amsmath}
\usepackage{stfloats}
\usepackage{graphicx}
\usepackage{subfigure}
\usepackage{epsfig,epsf,color,balance,cite}
\usepackage{url}
\usepackage{bm}
\usepackage{mathrsfs}
\usepackage{amsthm}

\usepackage[colorlinks]{hyperref}
\usepackage[applemac]{inputenc}
\usepackage[a4paper,left=1.4cm,right=1.4cm, top=1.8cm,bottom=1.8cm,
includehead,headheight=0pt,headsep=0pt]{geometry}

\newtheorem{lemma}{Lemma}
\newtheorem{remark}{Remark}
\usepackage{algorithm}
\usepackage{algpseudocode}
\usepackage{tensor}
\usepackage{graphicx}
\usepackage{epstopdf}
\IEEEoverridecommandlockouts
	
\begin{document}

\title{\LARGE Channel Estimation for FAS-assisted Multiuser mmWave Systems}
\vspace{-1mm}

\author{
	\IEEEauthorblockN{Hao Xu, \emph{Member, IEEE}\IEEEauthorrefmark{0},
		Gui Zhou, \emph{Member, IEEE}\IEEEauthorrefmark{0},     
		Kai-Kit Wong, \emph{Fellow, IEEE}\IEEEauthorrefmark{0},\\
		Wee Kiat New, \emph{Member, IEEE}\IEEEauthorrefmark{0},
		Chao Wang, \emph{Member, IEEE}\IEEEauthorrefmark{0},
		Chan-Byoung Chae, \emph{Fellow, IEEE}\IEEEauthorrefmark{0},\\
		Ross Murch, \emph{Fellow, IEEE}\IEEEauthorrefmark{0},
		Shi Jin, \emph{Fellow, IEEE}\IEEEauthorrefmark{0},
		and Yangyang Zhang
	}
\thanks{The work of H. Xu, K. K. Wong, and W. K. New is supported in part by the European Union's Horizon 2020 Research and Innovation Programme under MSCA Grant No. 101024636, and in part by the Engineering and Physical Sciences Research Council (EPSRC) under grant EP/W026813/1. The work of C. B. Chae is supported by the Institute of Information and Communication Technology Promotion (IITP) grant funded by the Ministry of Science and ICT (MSIT), Korea (No. 2021-0-02208, No. 2021-0-00486).}
\thanks{H. Xu, K. K. Wong, and W. K. New are with the Department of Electronic and Electrical Engineering, University College London, London WC1E7JE, United Kingdom. K. Wong is also affiliated with Yonsei Frontier Laboratory, Yonsei University, Seoul, 03722, Korea (e-mail: $\rm \{hao.xu, kai\text{-}kit.wong, a.new \}@ucl.ac.uk$). G. Zhou is with the Institute for Digital Communications, Friedrich-Alexander-University Erlangen-N{\"u}rnberg (FAU), 91054 Erlangen, Germany. C. Wang is with the Integrated Service Networks Laboratory, Xidian University, Xi'an 710071, China. C. B. Chae is with School of Integrated Technology, Yonsei University, Seoul, 03722, Korea. R. Murch is with the Department of Electronic and Computer Engineering and Institute for Advanced Study (IAS), Hong Kong University of Science and Technology, Clear Water Bay, Hong Kong SAR, China. S. Jin is with the Frontiers Science Center for Mobile Information Communication and Security, Southeast University, Nanjing, China. Y. Zhang is with Kuang-Chi Science Limited, Hong Kong SAR, China.}
\thanks{Corresponding authors: Gui Zhou and Kai-Kit Wong.}
\vspace{-2 em}
}

\maketitle

\vspace{-1mm}
\begin{abstract}
This letter investigates the challenge of channel estimation in a multiuser millimeter-wave (mmWave) time-division duplexing (TDD) system. In this system, the base station (BS) uses a multi-antenna uniform linear array (ULA), while each mobile user has a fluid antenna system (FAS). Accurate channel state information (CSI) plays a crucial role in the precise placement of antennas in FAS.
To tackle this issue, we propose a low-sample-size sparse channel reconstruction (L3SCR) method, capitalizing on the sparse propagation paths  of mmWave channels. 
Simulation results show that our method obtains precise CSI with minimal hardware switching and pilot overhead. As a result, the system sum-rate approaches the upper bound with perfect CSI.
\end{abstract}


\IEEEpeerreviewmaketitle

\vspace{-3mm}
\section{Introduction}\label{section1}
Recently, fluid antenna system (FAS) has emerged as a new technology to obtain spatial diversity \cite{wong2023fluid}. FAS refers to any software-controllable fluidic conductive structure, reconfigurable radio-frequency (RF)-pixels, or movable mechanical antenna structure that can change its shape and position to reconfigure the gain, radiation pattern, and other characteristics. By switching the antenna's position, a transmitter or receiver is able to access the ups and downs of the fading channel, providing additional degrees of freedom and significant communication gains. In \cite{wong2021fluid}, the outage probability of a point-to-point FAS-assisted system was first investigated, and the effect of the antenna size and port number was analyzed. FAS has also been proven to be effective in supporting multiple access and even massive connectivity. In \cite{wong2022fluid} and \cite{wong2023slow}, it was shown that by placing an antenna at the port where all interfering signals fade deeply, the interference at a user can be greatly reduced, making FAS-assisted multiple access possible.

Current studies on FAS mainly focus on the analysis and optimization of its communication performance, which relies heavily on the acquisition of channel state information (CSI). In FAS, the antenna can often change its position continuously in a given area, which means that to realize its full potential, the CSI of any position or a large number of preset ports at a FAS, has to be known, and conventional channel estimation schemes developed for fixed-antenna systems are no longer suitable. 
To date, little research has been done to estimate the CSI for FAS. In \cite{ma2023compressed}, channel estimation for a point-to-point communication system, where both the transmitter and receiver use a planar FAS,\footnote{We note that the term `movable antenna system' was used in \cite{ma2023compressed} instead. However, the term `FAS' is preferred as it includes both movable and non-movable position-flexible antennas such as on-off switching pixels \cite{Murch-2022}.} was studied. However, as shown by the results in \cite{ma2023compressed}, each antenna has to move over $256$ positions (the planar FAS has $400$ ports in total) to estimate the channel, which inevitably leads to quite high hardware switching and pilot overhead. Therefore, it is of great importance to design low-cost channel estimation schemes for FAS.

In this letter, we address the channel estimation problem for a multiuser millimeter-wave (mmWave) time-division duplexing (TDD) system, in which the base station (BS) uses a fixed multi-antenna uniform linear array (ULA) while each mobile user is equipped with a linear FAS. As a benchmark, we first show how to obtain the full CSI using the traditional least squares (LS) method. 
Utilizing the fact that mmWave channels have a sparse-scattering property \cite{zhou2022channel}, we then propose a low-sample-size sparse channel reconstruction (L3SCR) method, which works in three steps. In the first step, the antennas of the users switch over only a few estimating locations (ELs) and transmit orthogonal pilots at each EL, from which an LS estimation of the channel matrix with reduced dimension is obtained. In the second step, using a compressed-sensing based method, the sparse parameters, including the number of spatial paths, angles of arrival (AoAs), angles of departure (AoDs), and path gains, are extracted at the BS. Finally, based on the planar-wave geometric model and estimated sparse parameters, the complete channel matrix (between all ports and the BS) is reconstructed. Simulation results show that using the L3SCR method, accurate CSI information can be obtained with quite low hardware switching and pilot overhead. 

\vspace{-3mm}
\section{System Model}\label{GV_MAC-WT}
\begin{figure}[]
\centering
\includegraphics[scale=0.25]{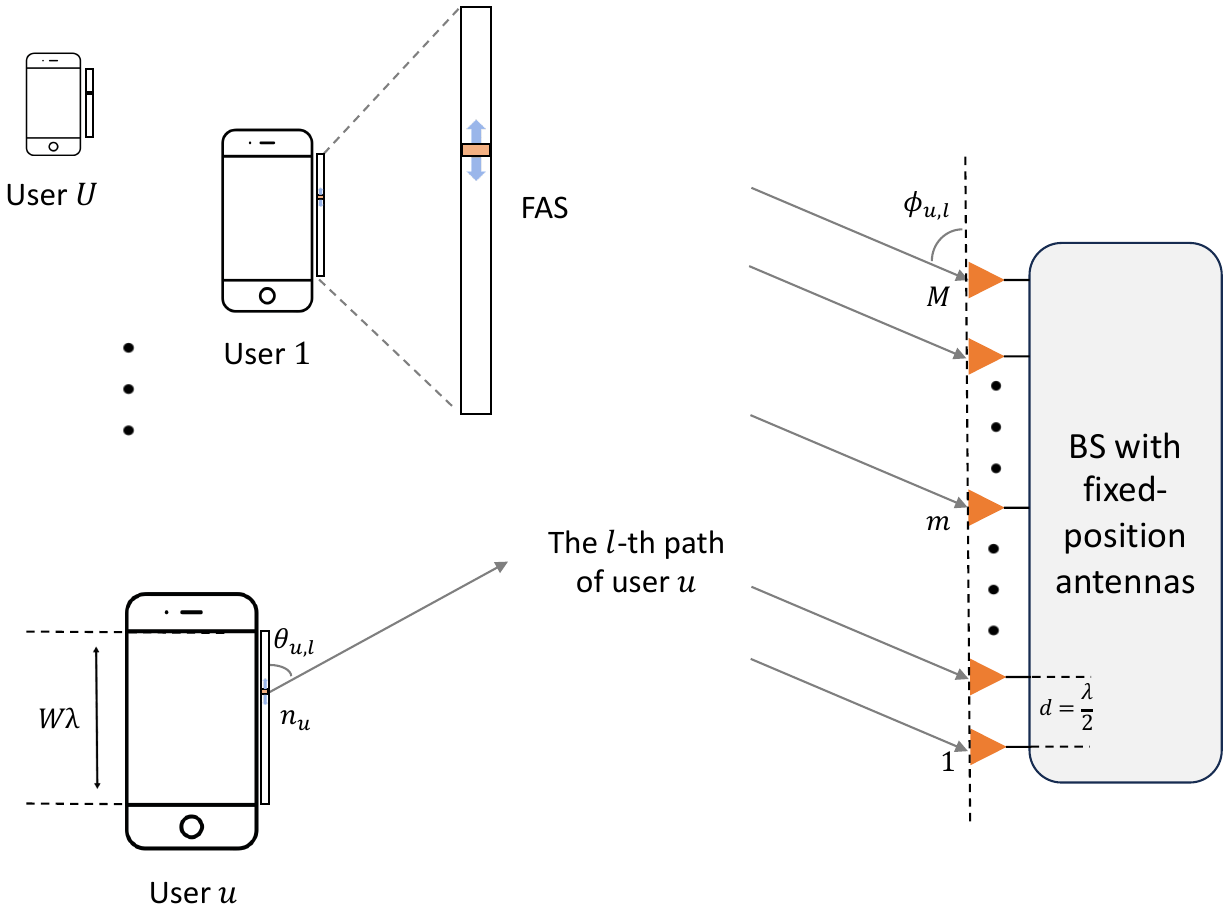}
\vspace{-1 em}
\caption{Illustration of a FAS-assisted uplink system.}\label{channel_model}
\vspace{-1.5 em}
\end{figure}

As shown in Fig.~\ref{channel_model}, we consider a narrow-band mmWave uplink system where $U$ users simultaneously communicate with the BS. Each user is equipped with a linear FAS of size $W \lambda$ and the fluid antenna can be instantly switched to one of the $N$ evenly distributed  ports. The BS is equipped with an $M$-antenna ULA with antenna spacing $d = \frac{\lambda}{2}$, where $\lambda$ is the carrier wavelength.\footnote{The proposed channel estimation method can be generalized to the case where the BS also applies FAS. In this case, the antenna at the BS also has to be switched among different ports to collect pilot signals at different positions.}
Let $n_u$ denote the port used by user $u$ for uplink transmission.
Then the received signal at the BS is 
\begin{equation}\label{yx}
\bm y = \sum_{u = 1}^U \bm g_{u, n_u} x_u + \bm z,
\end{equation}
where $x_u \sim {\cal {CN}} (0, p_u)$ is the data signal of user~$u$, $p_u$ is the transmit power, $\bm g_{u, n_u} \in {\mathbb C}^{M \times 1}$ is the channel vector from the $n_u$-th port of user $u$ to the BS, and $\bm z \sim {\cal {CN}} (\bm 0, \bm I_M)$ is the additive noise. The BS applies a linear receiver $\bm w_{u, n_u}$ to detect the signal of user $u$ when its antenna stops at port $n_u$. Then the sum-rate can be written as
\begin{equation}\label{rate_R}
R \!=\! \sum_{u = 1}^U \tau \log \!\Bigg(\! 1 \!+\! \frac{p_u \left| \bm w_{u, n_u}^H \bm g_{u, n_u} \right|^2 }{\sum\limits_{u' \neq u} p_{u'} \!\!\left| \bm w_{u, n_u}^H \bm g_{u', n_{u'}} \right|^2 \!+\! \left\| \bm w_{u, n_u} \right\|^2 } \!\Bigg),\!\!
\end{equation}
where $\tau$ is the loss from pilot signaling. The sum-rate can be maximized by choosing a `good' $n_u$ in $\{1, \dots, N\}$. 
We consider a quasi-static block-fading channel, and aim to estimate $\bm G_u\triangleq [ \bm g_{u,1}, \dots,\bm g_{u,N} ], \forall u \in \{1, \dots, U\}$.

\vspace{-2mm}
\section{LS Estimation}\label{LS}
In this section, we adopt the classical LS method to estimate $\bm G_u$. All users' antennas switch synchronously over $N$ ports and transmit orthogonal pilot sequences $\bm Q = [\bm q_1, \dots, \bm q_U] \in {\mathbb C}^{U \times U}$, where $\bm q_u^H \bm q_u = 1$ and $\bm q_u^H \bm q_{u'} = 0, \forall u \neq u'$. At each port, a pilot sequence is transmitted $T$ times, i.e., $T$ subframes are used for estimation with each subframe containing $U$ time slots. In the $t_n$-th subframe, all antennas are located at the $n$-th port and the received pilot signal at the BS is
\begin{equation}\label{y_nt}
\bm Y_n (t_n) = \sum_{u = 1}^U \sqrt{p_u} \bm g_{u,n} \bm q_u^H + \bm Z_n (t_n),
\end{equation}
where $\bm Z_n (t_n) \in {\mathbb C}^{M \times U}$ is the noise matrix whose elements are independent and identically distributed (i.i.d.) and all follow ${\cal {CN}} (0, 1)$. By right multiplying $\frac{1}{T \sqrt{p_u}} \sum_{t_n = 1}^T \bm Y_n (t_n)$ by $\bm q_u$, the LS estimate of $\bm g_{u,n}$ can be obtained as
\begin{align}\label{y_nt_qu}
\!{\hat {\bm g}}_{u,n}^{\text {LS}} \!=\! \frac{1}{T\! \sqrt{p_u}} \!\sum_{t_n = 1}^T \!\bm Y_n (t_n) \bm q_u
\!=\! \bm g_{u,n} \!+\! \frac{1}{T\! \sqrt{p_u}} \!\sum_{t_n = 1}^T \!\bm Z_n (t_n) \bm q_u.\!\!\!
\end{align}
We stack (\ref{y_nt_qu}) for all $n \in \{1, \dots, N\}$ and obtain 
\begin{align}\label{Gu_LS}
{\hat {\bm G}}_{u, {\text {LS}}} = \left[ {\hat {\bm g}}_{u,1}^{\text {LS}}, \dots, {\hat {\bm g}}_{u,N}^{\text {LS}} \right] = \bm G_u + {\hat {\bm Z}}_u,
\end{align}
where ${\hat {\bm Z}}_u \!=\! \frac{1}{T \!\sqrt{p_u}} \!\!\left[ \sum_{t_1 \!=\! 1}^T \!\bm Z_1 (\!t_1\!) \bm q_u, \dots, \sum_{t_N \!=\! 1}^T \!\bm Z_N (\!t_N\!) \bm q_u \!\right] \!\in\! {\mathbb C}^{M \!\times\! N}$. The performance of LS is determined by $T$ and $p_u$. However, since $N$ is large, letting the antennas switch and measure the channel requires extremely high overhead. 

\vspace{-2mm}
\section{L3SCR}\label{L3SCR}
To reduce the hardware switching and pilot overhead, in this section, we take advantage of the sparse-scattering property of mmWave channels and propose the L3SCR method. Using this method, the antennas only need to switch and measure over a few ELs, and the angular and gain information of the channel can be  extracted, from which the full CSI is then reconstructed.

Assume that during the estimation period each fluid antenna stops at only $K \ll N$ preset ELs, which are uniformly distributed with adjacent spacing $\Delta$ satisfying $\Delta \leq W \lambda/(K - 1)$. Let $\bm H_u = \left[ \bm h_{u,1}, \dots, \bm h_{u,K} \right] \in {\mathbb C}^{M \times K}$ denote the channel matrix between the $K$ ELs of user~$u$ and the BS. Using the planar-wave geometric channel model that is typically adopted in mmWave systems \cite{akdeniz2014millimeter}, $\bm H_u$ can be modeled as
\begin{equation}\label{H}
\bm H_u = \sqrt{MK} \sum_{l = 1}^{L_u} \gamma_{u,l} \bm a_{u,{\text R}} (\phi_{u,l}) \bm a_{u,{\text T}}^H (\theta_{u,l}),
\end{equation}
where $L_u$ is the number of propagation paths between user $u$ and the BS, and $\gamma_{u,l}$ is the complex channel gain. Also, $\bm a_{u,{\text R}} (\phi_{u, l})$ and $\bm a_{u,{\text T}} (\theta_{u, l})$ are respectively the steering vectors at the receiver and transmitter sides, given by
\begin{equation}\label{a_RT}
\left\{\!\!\!\begin{aligned}
\bm a_{u,{\text R}} (\phi_{u, l}) \!& = \!\! \frac{1}{\sqrt{\!M}} \!\left[\! 1, e^{\!-\! j \frac{2 \pi}{\lambda} d \cos \phi_{u, l}}, \dots, e^{\!-\! j \frac{2 \pi}{\lambda} (M\!-\!1) d \cos \phi_{u, l}} \!\right]^{\!T}\!\!\!,\!\!\! \\
\bm a_{u,{\text T}} (\theta_{u, l}) \!& = \!\!\frac{1}{\sqrt{\!K}} \!\left[\! 1, e^{\!-\! j \frac{2 \pi}{\lambda} \Delta \cos \theta_{u, l}}, \dots, e^{\!-\! j \frac{2 \pi}{\lambda} (K\!-\!1) \Delta \cos \theta_{u, l}} \!\right]^{\!T}\!\!\!,
\end{aligned}\right.
\end{equation}
where $\phi_{u, l}, \theta_{u, l} \in [0, \pi]$ are respectively the AoA and AoD of the $l$-th path. Denote\footnote{The order of the propagation paths in (\ref{A_RT}) can be arbitrarily changed and this will affect neither the value of $\bm H_u$ nor the following estimation process. }
\begin{equation}\label{A_RT}
\left\{\begin{aligned}
\bm \varGamma_u & = {\text {diag}} \{ \gamma_{u,1}, \dots, \gamma_{u, L_u} \} \in {\mathbb C}^{L_u \times L_u},\\
\bm A_{u,{\text R}} & = \left[ \bm a_{u,{\text R}} (\phi_{u, 1}), \dots, \bm a_{u,{\text R}} (\phi_{u, L_u}) \right] \in {\mathbb C}^{M \times L_u},\\
\bm A_{u,{\text T}} & = \left[ \bm a_{u,{\text T}} (\theta_{u, 1}), \dots, \bm a_{u,{\text T}} (\theta_{u, L_u}) \right] \in {\mathbb C}^{K \times L_u},
\end{aligned}\right.
\end{equation}
based on which $\bm H_u$ in (\ref{H}) can be rewritten in matrix form as
\begin{equation}\label{H2}
\bm H_u = \sqrt{MK} \bm A_{u,{\text R}} \bm \varGamma_u \bm A_{u,{\text T}}^H.
\end{equation}
Our aim is to estimate the sparse parameters $L_u$, $\gamma_{u,l}$, $\phi_{u, l}$, and $\theta_{u, l}$ such that $\bm G_u$ can be reconstructed. All users transmit orthogonal pilot sequences at $K$ ELs, and first get the LS estimation of $\bm H_u$ using the technique in Section~\ref{LS}, i.e.,
\begin{equation}\label{Hu_LS}
{\hat {\bm H}}_{u, {\text {LS}}} = \bm H_u + {\hat {\bm N}}_u,
\end{equation}
where ${\hat {\bm N}}_u \!\in\! {\mathbb C}^{M \times K}$ is similarly defined as ${\hat {\bm Z}}_u$ in (\ref{Gu_LS}).
Then, we extract the sparse parameters from (\ref{Hu_LS}) and reconstruct $\bm G_u$.

\subsection{Estimation of Number of Paths and AoAs}
We estimate the number of paths and AoAs by first applying discrete Fourier transform (DFT) to obtain a coarse estimation and then using angular rotation to refine the result. In particular, let $\bm \varOmega  \in {\mathbb C}^{M \times M}$ denote the normalized DFT matrix, whose $(m, m')$-th element is given by
\begin{equation}\label{U_mm}
[\bm \varOmega]_{m, m'} = \frac{1}{\sqrt{M}} e^{-j \frac{2 \pi}{M} (m - 1) (m' - 1)}.
\end{equation}
Then, the DFT of ${\hat {\bm H}}_{u, {\text {LS}}}$ normalized by $\sqrt{MK}$ is found as
\begin{align}\label{Y_DFT}
{\hat {\bm H}}_{u, {\text {LS}}}^{\text {DFT}} & = \frac{1}{\sqrt{MK}} \bm \varOmega^H {\hat {\bm H}}_{u, {\text {LS}}}\nonumber\\
& = \bm \varOmega^H \bm A_{u,{\text R}} \bm \varGamma_u \bm A_{u,{\text T}}^H + \frac{1}{\sqrt{M K}} \bm \varOmega^H {\hat {\bm N}}_u.
\end{align}

\begin{lemma}\label{lemma1}
If $M \rightarrow + \infty$, $\bm \varOmega^H \bm A_{u,{\text R}}$ is a row sparse matrix with a full column rank. Only one element in each column of $\bm \varOmega^H \bm A_{u,{\text R}}$ is $1$ while all the others are $0$.
\end{lemma}

\begin{proof}
See Appendix~\ref{prove_lemma1}.
\end{proof}

From Lemma~\ref{lemma1}, if $M$ approaches infinity, $\bm \varOmega^H \bm A_{u,{\text R}} \bm \varGamma_u \bm A_{u,{\text T}}^H$ is a row sparse matrix with $L_u$ non-zero rows, each corresponding to one of the AoAs $\phi_{u, l}$. Let $m_l$ denote the $l$-th non-zero row of $\bm \varOmega^H \bm A_{u,{\text R}} \bm \varGamma_u \bm A_{u,{\text T}}^H$. From Appendix~\ref{prove_lemma1}, we know that for a given $\phi_{u, l}$, if $\phi_{u, l} \in [0, \frac{\pi}{2}]$, $m_l$ satisfies $\frac{m_l - 1}{M} \leq \frac{d}{\lambda}$, and if $\phi_{u, l} \in (\frac{\pi}{2}, \pi]$, $\frac{m_l - 1}{M} > \frac{d}{\lambda}$. Then, if $m_l$ is known, $\phi_{u, l}$ can be directly calculated based on (\ref{cond1}) and (\ref{cond2}) as
\begin{equation}\label{rough_esti_phi}
\phi_{u, l} =\left\{
\begin{array}{ll}
\arccos \frac{(m_l - 1) \lambda}{M d}, &~{\text {if}}~ \frac{m_l - 1}{M} \leq \frac{d}{\lambda}, \\
\arccos \frac{(m_l - 1 - M) \lambda}{M d}, &~{\text {if}}~ \frac{m_l - 1}{M} > \frac{d}{\lambda}. 
\end{array} \right.
\end{equation}
Note that $M$ is in practice finite and what we observe is not $\bm \varOmega^H \bm A_{u,{\text R}} \bm \varGamma_u \bm A_{u,{\text T}}^H$ but its noisy version ${\hat {\bm H}}_{u, {\text {LS}}}^{\text {DFT}}$ (see (\ref{Y_DFT})). When $M$ is large, ${\hat {\bm H}}_{u, {\text {LS}}}^{\text {DFT}}$ is an asymptotic row sparse matrix and its power is concentrated on a few rows while the remaining power is leaked to the nearby rows. To estimate the AoAs from ${\hat {\bm H}}_{u, {\text {LS}}}^{\text {DFT}}$, we calculate the sum power of each row and search for ${\hat L_u}$ obvious power peaks where ${\hat L_u}$ can be regarded as an estimate of the number of paths $L_u$. Let ${\hat m}_l$ denote the index of the $l$-th power peak. Then, a coarse estimation of the AoAs can be obtained from (\ref{rough_esti_phi}) by replacing $m_l$ with ${\hat m}_l$.

The above DFT-based estimation is coarse and its performance is limited by the resolution $\frac{1}{M}$. With the estimated number of paths ${\hat L_u}$ and power peak indices $\{{\hat m}_1, \dots, {\hat m}_{\hat L_u}\}$, we further improve the performance by employing the angular rotation operation to compensate the angular mismatch \cite{fan2018angle}. Define the angular rotation matrix as
\begin{equation}\label{rotation_mat}
\bm \varPsi = {\text {diag}} \left\{ 1, e^{j 2 \pi \psi}, \dots, e^{j 2 \pi (M - 1) \psi} \right\},
\end{equation}
where $\psi \in [ - \frac{1}{2M}, \frac{1}{2M} ]$ is the rotation parameter. Applying both the DFT and angular rotation matrices to $\frac{1}{\sqrt{MK}} {\hat {\bm H}}_{u, {\text {LS}}}$, we then obtain
\begin{align}\label{Y_DFT_rota}
{\hat {\bm H}}_{u, {\text {LS}}}^{{\text {DFT}}, {\text {ro}}} 
= \bm \varOmega^H \bm \varPsi^H \bm A_{u,{\text R}} \bm \varGamma_u \bm A_{u,{\text T}}^H + \frac{1}{\sqrt{M K}} \bm \varOmega^H \bm \varPsi^H {\hat {\bm N}}_u.
\end{align}
The $(m, l)$-th element of $\bm \varOmega^H \bm \varPsi^H \bm A_{u,{\text R}}$ can be expressed as
\begin{align}\label{Y_DFT_rota_ml}
& [ \bm \varOmega^H \bm \varPsi^H \bm a_{u,{\text R}} (\phi_{u, l}) ]_m\nonumber\\
&= \frac{1}{M} \sum_{m' = 1}^M e^{- j 2 \pi (m' - 1) \left(\frac{d}{\lambda} \cos \phi_{u, l} - \frac{m - M \psi - 1}{M}\right)}.
\end{align}
Comparing (\ref{Y_DFT_rota_ml}) with (\ref{UA_lm}), we observe that the power beam at index $m$ is rotated to $m - M \psi$, which can vary continuously in $[m - \frac{1}{2}, m + \frac{1}{2}]$ (since $\psi \in [ - \frac{1}{2M}, \frac{1}{2M} ]$) and corresponds to a new angle for each $\psi$. Then, the remaining issue is for a given ${\hat m}_l$, how to find the best $\psi$ such that the estimated angle after compensation best matches the real one. We do this by
\begin{equation}\label{mu_l}
\psi_l = \arg \max_{\psi \in \{ - \frac{1}{2M}, - \frac{1}{2M} + \epsilon, \dots, \frac{1}{2M}\}} \parallel [ {\hat {\bm H}}_{u, {\text {LS}}}^{{\text {DFT}}, {\text {ro}}} ]_{{\hat m}_l, :}\parallel^2,
\end{equation}
where $\epsilon$ is the step length for searching. Once $\psi_l$ is determined, using (\ref{Y_DFT_rota_ml}) and following similar analysis in Appendix~\ref{prove_lemma1}, we could obtain the estimation of $\phi_{u, l}$ as
\begin{equation}\label{accurate_esti_phi}
{\hat \phi}_{u, l} = \left\{
\begin{array}{ll}
\arccos \frac{({\hat m}_l - M \psi_l - 1) \lambda}{M d}, &~{\text {if}}~ \frac{{\hat m}_l - 1}{M} \leq \frac{d}{\lambda},\\
\arccos \frac{({\hat m}_l - M \psi_l - 1 - M) \lambda}{M d}, &~{\text {if}}~ \frac{{\hat m}_l - 1}{M} > \frac{d}{\lambda}.
\end{array} \right.
\end{equation}
As such, ${\hat {\bm A}}_{u,{\text R}} = \left[ \bm a_{u,{\text R}} ({\hat \phi}_{u, 1}), \dots, \bm a_{u,{\text R}} ({\hat \phi}_{u, \hat L_u}) \right] \in {\mathbb C}^{M \times {\hat L_u}}$.

\vspace{-3mm}
\subsection{Estimation of AoDs and Channel Gains}
Assume that the estimation of the number of paths and AoAs is accurate.\footnote{If this is not true, the proposed scheme can still be applied. However, its performance may be compromised by errors in the number of paths and AoAs.} Using ${\hat {\bm A}}_{u,{\text R}}$, we project ${\hat {\bm H}}_{u, {\text {LS}}}^H$ onto the AoA steering matrix subspace as
\begin{align}\label{approx1}
& \frac{1}{\sqrt{M K}} {\hat {\bm H}}_{u, {\text {LS}}}^H {\hat {\bm A}}_{u,{\text R}} \!=\! \bm A_{u,{\text T}} \bm \varGamma_u ^H \bm A_{u,{\text R}}^H {\hat {\bm A}}_{u,{\text R}} \!+\! \frac{1}{\sqrt{M K}} {\hat {\bm N}}_u^H {\hat {\bm A}}_{u,{\text R}} \nonumber\\
& \approx \bm A_{u,{\text T}} \bm \varGamma_u ^H + \bm V_u \nonumber\\
& = \!\left[\! \gamma_{u,1}^* \bm a_{u,{\text T}} (\!\theta_{u, 1}\!) \!+\! \bm v_{u, 1}, \dots, \gamma_{u, L_u}^* \bm a_{u,{\text T}} (\!\theta_{u, L_u}\!) \!+\! \bm v_{u, L_u} \right]\!,\!\!\!\!
\end{align}
where the approximation holds since $\bm A_{u,{\text R}}^H {\hat {\bm A}}_{u,{\text R}} \approx \bm I_{L_u}$ when $M$ is large, $\bm V_u = \frac{1}{\sqrt{M K}} {\hat {\bm N}}_u^H {\hat {\bm A}}_{u,{\text R}}$, and $\bm v_{u, l}$ is the $l$-th column of $\bm V_u$. It is obvious that each column of (\ref{approx1}) contains the AoD and gain information of only one path. Hence, we are faced with a $1$-sparse reconstruction problem, and thus apply low-complexity matched filters to estimate the AoDs and gains. Define the following dictionary matrix
\begin{equation}\label{dict_matr}
\bm D \!=\! \left[\! \bm a_{u,{\text T}} (0), \bm a_{u,{\text T}} \!\left(\! \frac{1}{C} \pi \!\right), \dots, \bm a_{u,{\text T}} \!\left(\! \frac{C \!-\! 1}{C} \pi \!\right) \!\right] \!\!\in\! {\mathbb C}^{K \!\times\! C},\!\!
\end{equation}
where $C$ is the size of the dictionary, and $\bm a_{u,{\text T}} \left( \frac{c - 1}{C} \pi \right)$ is the array steering vector with angle $\frac{c - 1}{C} \pi$ and can be calculated from (\ref{a_RT}). Applying $\bm D$ to (\ref{approx1}), we get
\begin{align}\label{approx2}
& \frac{1}{\sqrt{MK}} \bm D^H {\hat {\bm H}}_{u, {\text {LS}}}^H {\hat {\bm A}}_{u,{\text R}}  \!\approx\! \left[ \gamma_{u,1}^* \bm D^H \bm a_{u,{\text T}} (\theta_{u, 1}) \!+\! \bm D^H \bm v_{u, 1}, \dots, \right. \nonumber\\
& \quad\quad\quad\quad\quad\quad\quad \left. \gamma_{u, L_u}^* \bm D^H \bm a_{u,{\text T}} (\theta_{u, L_u}) \!+\! \bm D^H \bm v_{u, L_u} \right].\!\!
\end{align}
The $c$-th element of $\gamma_{u,l}^* \bm D^H \bm a_{u,{\text T}} (\theta_{u, l})$ can be calculated as
\begin{align}\label{D_at_lc}
& \left[ \gamma_{u,l}^* \bm D^H \bm a_{u,{\text T}} (\theta_{u, l}) \right]_c = \gamma_{u,l}^* \bm a_{u,{\text T}} \left( \frac{c - 1}{C} \pi \right)^H \bm a_{u,{\text T}} (\theta_{u, l}) \nonumber\\
& = \frac{\gamma_{u,l}^*}{K} \sum_{k = 1}^K e^{j \frac{2 \pi}{\lambda} (k-1) \Delta \left[ \cos \left( \frac{c - 1}{C} \pi \right) - \cos \theta_{u, l} \right]}.
\end{align}
Obviously, the modulus of (\ref{D_at_lc}) satisfies
\begin{align}\label{abs_D_at_lc}
\left| \left[ \gamma_{u,l}^* \bm D^H \bm a_{u,{\text T}} (\theta_{u, l}) \right]_c \right| \leq | \gamma_{u,l}^* |.
\end{align}
When $C$ is large enough, we can always find an integer $c_l \in \{ 1, \dots, C\}$ such that $\cos \left( \frac{c_l - 1}{C} \pi \right) - \cos \theta_{u, l} = 0$ and $\left[ \gamma_{u,l}^* \bm D^H \bm a_{u,{\text T}} (\theta_{u, l}) \right]_c = \gamma_{u,l}^*$. In this case, (\ref{abs_D_at_lc}) holds with equality, indicating that if depicting the power of the elements of $\gamma_{u,l}^* \bm D^H \bm a_{u,{\text T}} (\theta_{u, l}) + \bm D^H \bm v_{u, l}$ over their indices, we see a peak at 
\begin{equation}\label{c_l}
c_l = \arg \max_{c \in \{ 1, \dots, C\}} \left| \left[ \gamma_{u,l}^* \bm D^H \bm a_{u,{\text T}} (\theta_{u, l}) + \bm D^H \bm v_{u, l} \right]_c \right|.
\end{equation}
Based on this observation, $\theta_{u, l}$ and $\gamma_{u,l}$ can be estimated as
\begin{align}
{\hat \theta}_{u, l} & = \frac{c_l - 1}{C} \pi, \\
{\hat \gamma}_{u, l} & = \left[ \gamma_{u,l}^* \bm D^H \bm a_{u,{\text T}} (\theta_{u, l}) + \bm D^H \bm v_{u, l} \right]_{c_l}^*.
\end{align}

\begin{remark}\label{remark1}
From (\ref{D_at_lc}), we know that if $K = 1$, then $\left[ \gamma_{u,l}^* \bm D^H \bm a_{u,{\text T}} (\theta_{u, l}) \right]_c = \gamma_{u,l}^*, \forall c \in \{1, \dots, C\}$. Obviously, the AoDs cannot be estimated in this case. Therefore, the L3SCR scheme works only if $K \geq 2$, i.e., each fluid antenna must transmit pilot from at least two different ELs.
\end{remark}

\begin{remark}\label{remark2}
It should be noted that the inequality (\ref{abs_D_at_lc}) holds with equality not only when $\cos \left( \frac{c - 1}{C} \pi \right) - \cos \theta_{u, l} = 0$, but also whenever $\frac{\Delta}{\lambda} \left[ \cos \left( \frac{c - 1}{C} \pi \right) - \cos \theta_{u, l} \right]$ is an integer (no matter positive or not).	It can be easily verified that if $\Delta \geq \lambda$, for any $\theta_{u, l} \in [0, \pi]$, there are multiple $c \in \{ 1, \dots, C\}$ such that $\frac{\Delta}{\lambda} \left[ \cos \left( \frac{c - 1}{C} \pi \right) - \cos \theta_{u, l} \right]$ is an integer and thus (\ref{abs_D_at_lc}) holds. In order to avoid angular mismatch, we restrict $\Delta < \lambda$, i.e., the switching step of the fluid antenna should be less than the wavelength in the estimation.
\end{remark}

\subsection{Channel Reconstruction}
Once the number of paths, AoAs, AoDs, and channel gains are estimated, we can reconstruct $\bm G_u$ based on the planar-wave geometric model. In particular, we have
\begin{equation}\label{G_hat}
{\hat {\bm G}}_{u, {\text {L3SCR}}} = \sqrt{M N} \sum_{l = 1}^{{\hat L}_u} {\hat \gamma}_{u,l} \bm a_{u,{\text R}} ({\hat \phi}_{u,l}) {\hat {\bm a}}_{u,{\text T}}^H ({\hat \theta}_{u,l}),
\end{equation}
where $\bm a_{u,{\text R}} ({\hat \phi}_{u,l})$ can be calculated based on (\ref{a_RT}) and
\begin{equation}\label{a_T_hat}
\!\!{\hat {\bm a}}_{u,{\text T}} ({\hat \theta}_{u, l}) \!=\! \frac{1}{\sqrt{\!N}} \!\!\left[ 1, e^{\!-\! j 2 \pi \frac{W}{N\!-\!1} \!\cos {\hat \theta}_{u, l}}, \dots, e^{\!-\! j 2 \pi W \!\cos {\hat \theta}_{u, l}} \right]^{\! T}\!\!\!.\!\!\!
\end{equation}

\subsection{Analysis of Pilot Overhead and Computational Complexity}

Using the L3SCR scheme, the antenna of each user switches over $K$ ELs, and at each EL, a pilot of length $U$ is transmitted $T$ times. As a result, the total pilot overhead is $KTU$. In contrast, the LS method requires $N$ hardware switches and $NTU$ pilot overhead. As we will show in the next section, an accurate estimation of the full CSI can be obtained when $K$ and $T$ are, respectively, $6$ and $1$. Therefore, the L3SCR scheme requires very low hardware switching and pilot overhead.

To facilitate the analysis, we assume equal number of paths for all users, i.e., $L_u = L, \forall u \in \{1, \dots, U\}$.
The complexity of estimating the number of paths and AoAs mainly stems from the angle rotation operation (\ref{mu_l}), which involves $\frac{1}{M \epsilon}$ matrix multiplications in (\ref{Y_DFT_rota}), each with a complexity of ${\cal O} (M^3)$.
Hence, estimating the number of paths and AoAs for each user requires a complexity of ${\cal O} (M^2 \frac{1}{\epsilon})$.
The complexity of estimating the AoDs and channel gains mainly stems from (\ref{c_l}), which requires $C$ matrix multiplications in (\ref{approx2}), each with a complexity of ${\cal O} (C K L)$.
The overall complexity of the L3SCR scheme is thus ${\cal O} (U M^2 \frac{1}{\epsilon} + U C^2 K L)$.


\section{Simulation Results}\label{simul}
\begin{figure}
\centering
\includegraphics[scale=0.42]{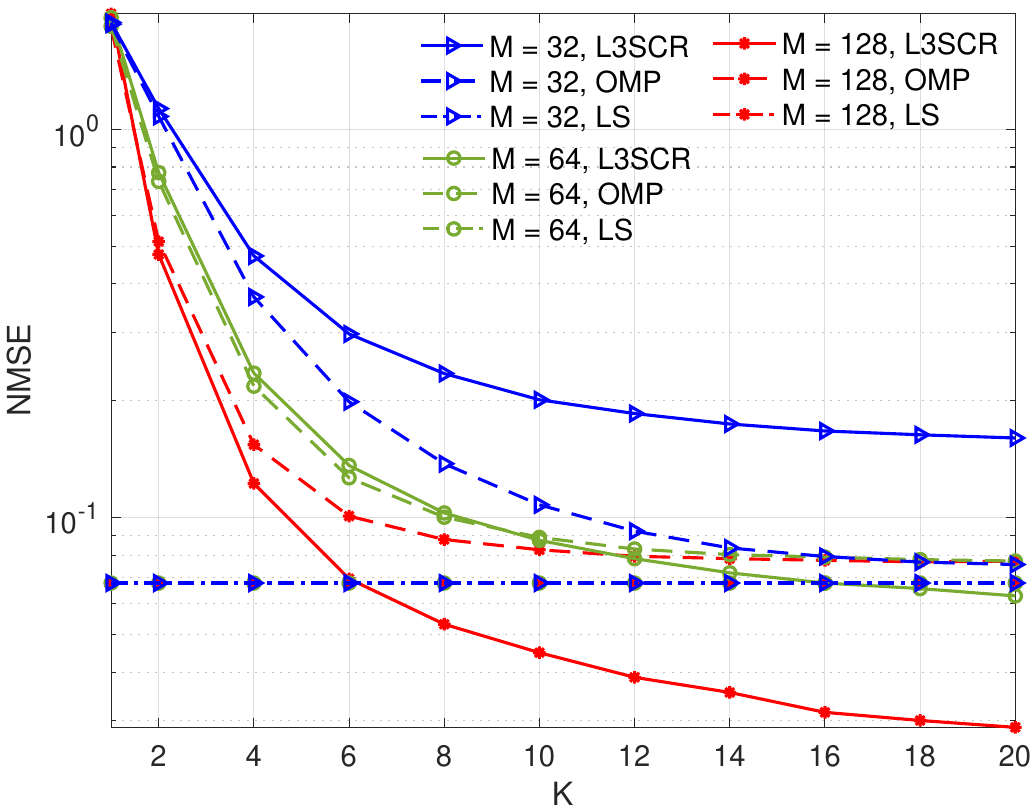}
\vspace{-1 em}
\caption{NMSE versus $K$ with $\rho = 10$~dB and $T = 1$.}\label{NMSE_VS_KM}
\vspace{-1.5 em}
\end{figure}

In this section, we evaluate the performance of the proposed L3SCR scheme by simulations. The uplink carrier frequency is set to $28$ GHz. For convenience, we assume equal maximum power constraint and equal number of paths for all users, i.e., $p_u = p, L_u = L, \forall u \in \{1, \dots, U\}$, and define the signal-to-noise ratio (SNR) as $\rho = 10 \lg p$ (dB). We set $U = 2$, $W = 10$, $N = 50$, $\Delta = \frac{\lambda}{2}$, $L = 5$, $\epsilon = \frac{1}{400 M}$, and $C = 10^4$. Let $S$ denote the number of symbols in a coherence block, which is mainly determined by the coherence bandwidth and coherence time \cite{bjornson2016massive}. Here we set $S = 500$. Since the L3SCR and LS schemes respectively require $KTU$ and $NTU$ pilot overhead, their corresponding loss factors in (\ref{rate_R}) are given by
\begin{equation}\label{tau}
	\tau_{\text {L3SCR}} = 1 - \frac{KTU}{S}, ~ \tau_{\text {LS}} = 1 - \frac{NTU}{S}.
\end{equation}
Besides the LS method, we also compare the proposed L3SCR scheme with the orthogonal matching pursuit (OMP) algorithm in \cite{lee2016channel}. In particular, we use the OMP algorithm to estimate the sparse parameters from (\ref{Hu_LS}) and then reconstruct $\bm G_u$.
All simulation results are obtained by averaging over $1000$ channel realizations. We investigate two performance metrics. The first one is the normalized mean square error (NMSE)
\begin{equation}\label{NMSE}
{\text {NMSE}} ({\hat {\bm G}}_u) = {\mathbb E} \left[ \frac{\sum_{u = 1}^U \parallel \!\!\bm G_u - {\hat {\bm G}}_u \!\!\parallel_F^2}{\sum_{u = 1}^U \parallel\!\! \bm G_u \!\!\parallel_F^2} \right],
\end{equation}
where the expectation is taken over different channel realizations, and ${\hat {\bm G}}_u$ can be any of ${\hat {\bm G}}_{u, {\text {L3SCR}}}$, ${\hat {\bm G}}_{u, {\text {LS}}}$, or ${\hat {\bm G}}_{u, {\text {OMP}}}$. The second metric is the average system rate $R$, which can be calculated from (\ref{rate_R}). Here we assume that the BS adopts the maximum ratio combining method for signal detection. In addition, we look for the optimal ports for the users to maximize $R$ by employing the exhaustive searching method.

\begin{figure}
\centering
\includegraphics[scale=0.42]{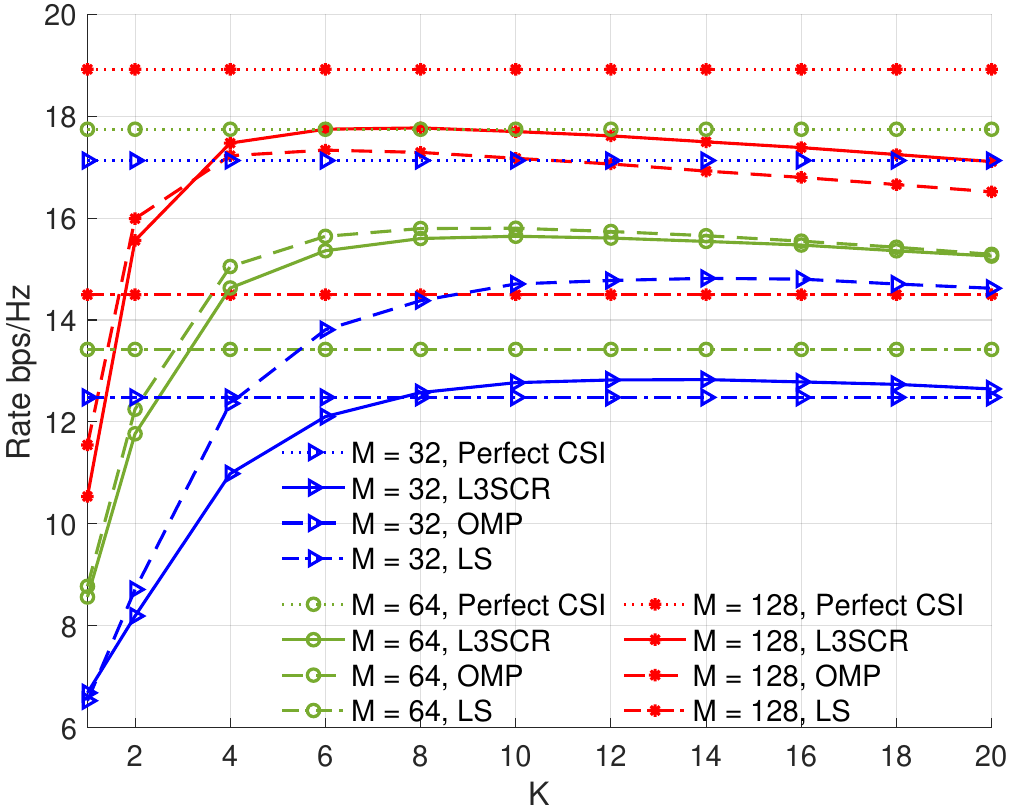}
\vspace{-1 em}
\caption{Average sum rate versus $K$ with $\rho = 10$~dB and $T = 1$.}\label{rate_VS_KM}
\vspace{-1.5 em}
\end{figure}

\begin{figure}
\centering
\includegraphics[scale=0.42]{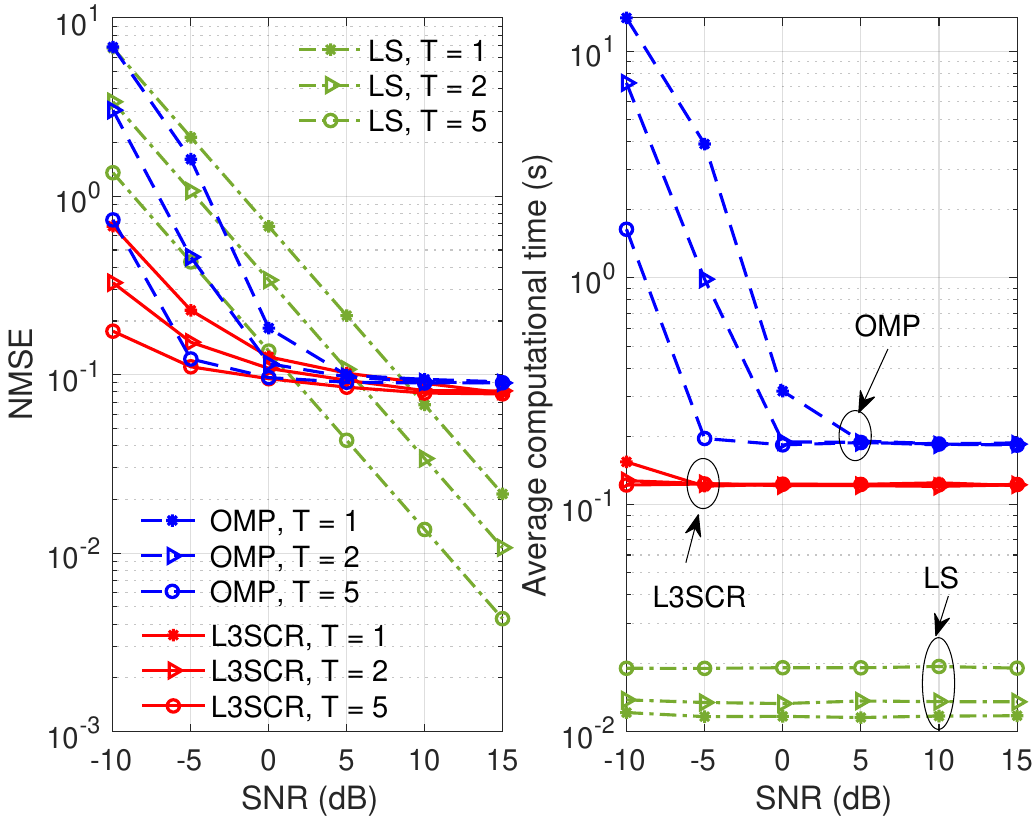}
\vspace{-1 em}
\caption{NMSE and computational time with $M = 64$ and $K = 10$.}\label{NMSE_time_VS_PT}
\vspace{-1.5 em}
\end{figure}

In Fig.~\ref{NMSE_VS_KM} and Fig.~\ref{rate_VS_KM}, the effect of $M$ and $K$ is investigated, from which several observations can be made. First, as $M$ increases, both ${\text {NMSE}} ({\hat {\bm G}}_{u, {\text {L3SCR}}})$ and ${\text {NMSE}} ({\hat {\bm G}}_{u, {\text {OMP}}})$ notably drop, leading to great improvement in the sum rate. When $M$ is small, the OMP algorithm outperforms the proposed L3SCR scheme in terms of both NMSE and sum rate, while when $M$ is large, the situation reverses. This is because the L3SCR scheme separately estimates the AoAs and AoDs. The estimation error in the first step will therefore be passed to the second step. In contrast, the OMP algorithm in \cite{lee2016channel} jointly estimates the AoAs and AoDs. Therefore, if $M$ is small, the OMP algorithm scheme can be applied to improve the estimation accuracy. Second, it is observed that when $K$ increases, the NMSE of both the L3SCR and OMP schemes decreases monotonically, whereas the sum rate first increases and then decreases. This is because while increasing $K$ can effectively improve estimation accuracy, it also increases pilot overhead. Moreover, we see that ${\text {NMSE}} ({\hat {\bm G}}_{u, {\text {LS}}})$ remains unchanged with $K$ and $M$. We will explain this later in Fig.~\ref{NMSE_time_VS_PT}. In many configurations, the LS method outperforms the L3SCR and OMP schemes in terms of NMSE. However, it is important to recognize that the LS method requires $NTU$ pilot overhead, resulting in lower sum rate (see Fig.~\ref{rate_VS_KM}).

Fig.~\ref{NMSE_time_VS_PT} depicts the NMSE and computational time required for one channel realization versus $\rho$ and $T$. It can be observed that $\lg {\text {NMSE}} ({\hat {\bm G}}_{u, {\text {LS}}})$ decreases linearly with $\rho$, while $\lg {\text {NMSE}} ({\hat {\bm G}}_{u, {\text {L3SCR}}})$ and $\lg {\text {NMSE}} ({\hat {\bm G}}_{u, {\text {OMP}}})$ reduce at the beginning and then saturate. Using (\ref{Gu_LS}), (\ref{NMSE}), and the fact that $p = 10^{\rho / 10}$, $\lg {\text {NMSE}} ({\hat {\bm G}}_{u, {\text {LS}}})$ can be calculated as
\begin{equation}\label{NMSE_LS}
\lg {\mathbb E} \left[ \frac{\sum_{u = 1}^U \left\| \sum_{t = 1}^T {\hat {\bm Z}}_u (t) \right\|_F^2}{\sum_{u = 1}^U \left\| \bm G_u \right\|_F^2} \right] - 2 \lg T - \frac{\rho}{10},
\end{equation}
where the expectation is seen as a constant. (\ref{NMSE_LS}) shows that $\lg {\text {NMSE}} ({\hat {\bm G}}_{u, {\text {LS}}})$ is determined only by $\rho$ and $T$, and decreases linearly with $\rho$. This also explains why it remains unchanged with $K$ and $M$ in Fig.~\ref{NMSE_VS_KM}. In contrast, the performance of L3SCR and OMP is affected not only by $\rho$ and $T$, but also by $M$ and $K$ (see Fig.~\ref{NMSE_VS_KM}). In addition, we see that the computational time of both L3SCR and OMP decreases with $\rho$, and the OMP algorithm requires much higher computational complexity. This is because when $\rho$ is small, more iterations are needed for the algorithms to converge, and different from the L3SCR scheme, the OMP algorithm needs to compute matrix inversions, which is computationally intensive.

\section{Conclusions}\label{section6}
This letter devised a novel channel estimation scheme for a multiuser mmWave system, wherein mobile users employ FAS to establish communication with the BS. Leveraging the channel sparsity, our scheme mandates each fluid antenna to selectively switch and measure the channel across a limited set of ELs. This approach enables us to reconstruct full CSI with minimal hardware switching and pilot overhead. Our results substantiated the efficacy of our proposed scheme, both in terms of NMSE and the system sum-rate performance.
\appendices

\section{Proof of Lemma~\ref{lemma1}}\label{prove_lemma1}
Denote
\begin{equation}\label{UA_l}
\bm \varOmega^H \bm A_{u,{\text R}} = \left[ \bm \varOmega^H \bm a_{u,{\text R}} (\phi_{u, 1}), \dots, \bm \varOmega^H \bm a_{u,{\text R}} (\phi_{u, L_u}) \right].
\end{equation}
Based on the definitions of $\bm a_{u,{\text R}} (\phi_{u, l})$ and $\bm \varOmega$ in (\ref{a_RT}) and (\ref{U_mm}), the $m$-th element of $\bm \varOmega^H \bm a_{u,{\text R}} (\phi_{u, l})$ can be written as
\begin{align}\label{UA_lm}
& [ \bm \varOmega^H \bm a_{u,{\text R}} (\phi_{u, l}) ]_m \nonumber\\
= & \frac{1}{M} \sum_{m' = 1}^M e^{j \frac{2 \pi}{M} (m' - 1) (m - 1)} e^{- j \frac{2 \pi}{\lambda} (m' - 1) d \cos \phi_{u, l}} \nonumber\\
= & \frac{1}{M} \sum_{m' = 1}^M e^{- j 2 \pi (m' - 1) \left(\frac{d}{\lambda} \cos \phi_{u, l} - \frac{m - 1}{M}\right)}.
\end{align}
We prove Lemma~\ref{lemma1} by separately discussing two cases with $\phi_{u, l} \in [0, \frac{\pi}{2}]$ and $\phi_{u, l} \in (\frac{\pi}{2}, \pi]$. 

First, if $\phi_{u, l} \in [0, \frac{\pi}{2}]$ and $M$ is large enough, since $\frac{d}{\lambda} = \frac{1}{2}$, there always exists an integer $m_l \in \{ 1, \dots, M\}$ such that
\begin{equation}\label{cond1}
\frac{d}{\lambda} \cos \phi_{u, l} - \frac{m_l - 1}{M} = 0.
\end{equation}
Then, it is known from (\ref{UA_lm}) that
\begin{equation}\label{m_1}
[ \bm \varOmega^H \bm a_{u,{\text R}} (\phi_{u, l}) ]_{m_l} = 1.
\end{equation}
For any $m \in \{ 1, \dots, M\} \setminus m_l$, it can always be expressed as $m = m_l + f$, where $f \in {\mathbb N}$ is non-zero. Using the sum formula of a geometric progression and (\ref{cond1}), (\ref{UA_lm}) can be rewritten as
\begin{align}\label{m_0}
[ \bm \varOmega^H \bm a_{u,{\text R}} (\phi_{u, l}) ]_m & = \frac{1}{M} \frac{1 - e^{- j 2 \pi M \left(\frac{d}{\lambda} \cos \phi_{u, l} - \frac{m - 1}{M}\right)}}{1 - e^{- j 2 \pi \left(\frac{d}{\lambda} \cos \phi_{u, l} - \frac{m - 1}{M}\right)}} \nonumber\\
& = \frac{1}{M} \frac{1 - e^{j 2 \pi f}}{1 - e^{- j 2 \pi \frac{f}{M} }} \nonumber\\
& = 0, \forall m \in \{ 1, \dots, M\} \setminus m_l.
\end{align}

If $\phi_{u, l} \in (\frac{\pi}{2}, \pi]$, we use the fact that $e^{- j 2 \pi (m' - 1) } = 1, \forall m' \in \{ 1, \dots, M\} \setminus m_l$ and rewrite (\ref{UA_lm}) as
\begin{align}\label{UA_lm3}
& [ \bm \varOmega^H \bm a_{u,{\text R}} (\phi_{u, l}) ]_m \nonumber\\
= & \frac{1}{M} \sum_{m' = 1}^M e^{- j 2 \pi (m' - 1) } e^{- j 2 \pi (m' - 1) \left(\frac{d}{\lambda} \cos \phi_{u, l} - \frac{m - 1}{M}\right)} \nonumber\\
= & \frac{1}{M} \sum_{m' = 1}^M e^{- j 2 \pi (m' - 1) \left(\frac{d}{\lambda} \cos \phi_{u, l} + 1 - \frac{m - 1}{M}\right)}.
\end{align}
Obviously, if $M$ is large enough, there always exists an integer $m_l \in \{ 1, \dots, M\}$ such that
\begin{equation}\label{cond2}
\frac{d}{\lambda} \cos \phi_{u, l} + 1 - \frac{m_l - 1}{M} = 0.
\end{equation}
Then, it can be proven similarly as the first case that (\ref{m_1}) and (\ref{m_0}) also hold. This completes the proof of Lemma~\ref{lemma1}.


\begin{thebibliography}{10}
\providecommand{\url}[1]{#1}
\csname url@samestyle\endcsname
\providecommand{\newblock}{\relax}
\providecommand{\bibinfo}[2]{#2}
\providecommand{\BIBentrySTDinterwordspacing}{\spaceskip=0pt\relax}
\providecommand{\BIBentryALTinterwordstretchfactor}{4}
\providecommand{\BIBentryALTinterwordspacing}{\spaceskip=\fontdimen2\font plus
\BIBentryALTinterwordstretchfactor\fontdimen3\font minus
  \fontdimen4\font\relax}
\providecommand{\BIBforeignlanguage}[2]{{%
\expandafter\ifx\csname l@#1\endcsname\relax
\typeout{** WARNING: IEEEtran.bst: No hyphenation pattern has been}%
\typeout{** loaded for the language `#1'. Using the pattern for}%
\typeout{** the default language instead.}%
\else
\language=\csname l@#1\endcsname
\fi
#2}}
\providecommand{\BIBdecl}{\relax}
\BIBdecl

\bibitem{wong2023fluid}
K.-K. Wong, W.~K. New, X.~Hao, K.-F. Tong, and C.-B. Chae, ``Fluid antenna system -- {Part I}: Preliminaries,'' \emph{IEEE Commun. Lett.}, vol.~27, no.~8, pp. 1919--1923, Aug. 2023.


\bibitem{wong2021fluid}
K.-K. Wong, A.~Shojaeifard, K.-F. Tong, and Y.~Zhang, ``Fluid antenna systems,'' \emph{IEEE Trans. Wireless Commun.}, vol.~20, no.~3, pp. 1950--1962, Mar. 2021.

\bibitem{wong2022fluid}
K.-K. Wong and K.-F. Tong, ``Fluid antenna multiple access,'' \emph{IEEE Trans. Wireless Commun.}, vol.~21, no.~7, pp. 1950--1962, Jul. 2022.

\bibitem{wong2023slow}
K.-K. Wong, D.~Morales-Jimenez, K.-F. Tong, and C.-B. Chae, ``Slow fluid antenna multiple access,'' \emph{IEEE Trans. Commun.}, vol.~71, no.~5, pp.  2831--2846, May 2022.


\bibitem{ma2023compressed}
W.~Ma, L.~Zhu, and R.~Zhang, ``Compressed sensing based channel estimation for movable antenna communications,'' \emph{IEEE Commun. Lett.}, 2023.

\bibitem{Murch-2022}
L. Jing, M. Li and R. Murch, ``Compact pattern reconfigurable pixel antenna with diagonal pixel connections,'' {\em IEEE Trans. Antennas \& Propag.}, vol. 70, no. 10, pp. 8951--8961, Oct. 2022.


\bibitem{zhou2022channel}
G.~Zhou, C.~Pan, H.~Ren, P.~Popovski, and A.~L. Swindlehurst, ``Channel estimation for {RIS}-aided multiuser millimeter-wave systems,'' \emph{IEEE Trans. Sig. Process.}, vol.~70, pp. 1478--1492, Mar. 2022.

\bibitem{akdeniz2014millimeter}
M.~R. Akdeniz {\em et al.}, ``Millimeter wave channel modeling and cellular capacity evaluation,'' \emph{IEEE J. Sel. Areas Commun.}, vol.~32, no.~6, pp. 1164--1179, Jun. 2014.

\bibitem{fan2018angle}
D.~Fan {\em et al.}, ``Angle domain channel estimation in hybrid millimeter wave massive {MIMO} systems,'' \emph{IEEE Trans. Wireless Commun.}, vol.~17, no.~12, pp. 8165--8179, Dec. 2018.

\bibitem{bjornson2016massive}
E.~Bj{\"o}rnson, E.~Larsson, and T.~Marzetta, ``Massive {MIMO}: Ten myths and one critical question,'' \emph{IEEE Commun. Mag.}, vol.~54, no.~2, pp. 114--123, Feb. 2016.

\bibitem{lee2016channel}
J.~Lee, G.~Gil, and Y.~Lee, ``Channel estimation via orthogonal matching pursuit for hybrid {MIMO} systems in millimeter wave communications,'' \emph{IEEE Trans. Commun.}, vol.~64, no.~6, pp. 2370--2386, Jun. 2016.
\end{thebibliography}

\end{document}